\documentclass[journal]{IEEEtran}
\usepackage{amssymb}
\usepackage[T1]{fontenc}
\usepackage{graphicx}
\usepackage{subfigure}
\usepackage{amsmath}
\usepackage{cite}
\usepackage{tabu}

\ifCLASSINFOpdf
\else
\fi
\begin{document}
%
\title{On the Scheduling Policy for Multi-process WNCS under Edge Computing \thanks{
	Y. Qiu, S. Wu and Y. Wang are with the Harbin Institute of Technology, Shenzhen, Guangdong, 518055, China. E-mail: 20s052037@stu.hit.edu.cn, hitwush@hit.edu.cn, wangy\_hit@foxmail.com}}%

%
%
%

\author{\IEEEauthorblockN{Yifei~Qiu, Shaohua~Wu, and Ying~Wang\\}}

\maketitle
\begin{abstract}
This paper considers a multi-process and multi-controller wireless networked control system (WNCS). There are $N$ independent linear time-invariant processes in the system plant which represent different kinds of physical processes. By considering the edge computing, the controllers are played by edge server and cloud server. Each process is measured by a sensor, and the status updates is sent to controller to generate the control command. The link delay of cloud server is longer than that of edge server. The processing time of status update depends on the characteristic of servers and processes. By taking into account such conditions, we mainly investigate how to choose the destination of status updates to minimize the system's average Mean Square Error (MSE), edge server or cloud server? To address this issue, we formulate an infinite horizon average cost Markov Decision Process (MDP) problem and obtain the optimal scheduling policy. The monotonicity of the value function in MDP is characterized and then used to show the threshold structure properties of the optimal scheduling policy. To overcome the curse of dimensionality, we propose a low-complexity suboptimal policy by using additive separable structure of value function. Furthermore, the processing preemption mechanism is considered to handle the status updates more flexible, and the consistency property is proved. On this basis, a numerical example is provided. The simulation results illustrate that the controller selection is related to the timeliness of process and show the performance of the suboptimal policy. We also find that the optimal policy will become a static policy in which the destination of status update is fixed when the wireless channel is error free.

\end{abstract}

\begin{IEEEkeywords}
WNCS, multi-process system, Edge computing, age of information, preemption, schelduling
\end{IEEEkeywords}

%
\IEEEpeerreviewmaketitle

\section{Introduction}
%
%
%
%
\IEEEPARstart{I}{n} a wireless networked control system (WNCS), to ensure the accuracy of control and the stability of design, it is necessary to describe the timeliness and error of the system quantitatively. There are two commonly used parameters: Age of information (AoI) and Mean Square Error (MSE). The concept of AoI was proposed in \cite{tsai2020unifying}, which can accurately quantify the timeliness of status information. Typically, the AoI is defined as the time elapsed since the most recently received status update was generated at the sensor. Unlike AoI, MSE is mainly used to measure the system's error, which can describe the change rates of different sources. It is widely used, especially in the control system, such as the multi-sensor control system in \cite{zang2020over}.

Based on AoI and MSE, the research under the background of WNCS has yielded many beneficial results. Some literature study the scheduling polices to optimize the system AoI or MSE. For example, the work in \cite{hsu2018age} studies wireless networks with multiple users under the problem of AoI minimization and develops a low-complexity transmission scheduling algorithm. The authors in \cite{jiang2018can} extend the multiple users system to incorporate stochastic packet arrivals and optimal packet management and proposes an index-prioritized random access policy by minimizing the AoI. The works in \cite{tsai2020unifying} and \cite{huang2020real} obtained the optimal actions in current state by designing the indicator based on latency and reliability, respectively. The authors in \cite{huang2020real} and \cite{wu2017optimal} study different lengths of source status and give the optimal policies in many cases by minimizing the system AoI. Based on \cite{wu2017optimal}, the work in \cite{wang2019preempt} investigates the preemption mechanism in channel and prove the structural properties of the optimal policy. The non-uniform status model was further extended in \cite{zhou2019minimum}, which is no longer limited to a single source and considers the situation of multiple processes with different status updates. The work in  \cite{wu2017optimalschduling} introduces an optimal policy considering the change rate of different sources by minimizing the system MSE.

In WNCS, smaller delay means more accurate control, but the existing scheduling policies can not reduce the inherent delay in uplink and downlink. Therefore, edge computing was proposed to decrease prolonged delays in some time-critical tasks. In some WNCS, a small set of special servers was deployed near the sensors. These special servers are called edge servers, which nearby sensors can reach via wireless connections.
In \cite{wang2020survey} and \cite{han2019ondisc}, edge computing is added to the sensor network. The authors in \cite{wang2020survey} introduce that the sensors can only full-offload the task, i.e., each status update can only be processed by one server, which can not be divided into multiple parts and handed over to multiple servers. The work in \cite{shiri2019massive} introduces a neural network for training the system model and considers the processing time on server. The work in \cite{ skarin2018towards } uses the existing cloud processing center to build an edge computing model, and the results show that the computing capacity is different between edge server and cloud server. The authors in \cite{han2019ondisc} propose a universal model which uses delay and computational capacity to distinguish different servers. The full task offloading without task migration in sensor network is also considered in \cite{han2019ondisc}.

Most existing works, e.g., [2]-[11], assume that the system has only one controller, and it takes the same time for different status updates to generate the control command. However, for the WNCS system containing edge computing, there are often multiple controllers in the system, and the controllers are mainly divided into edge servers and cloud servers. In addition, the time of generating control commands is also different.
The edge servers are less potent in both resources and computational ability than remote cloud servers. It take longer time to process state information and generate control commands than cloud servers \cite{shiri2019massive}. Moreover, to timely respond to time-critical tasks and improve system performance, the preemption mechanism commonly exists on the server, and tasks with higher priority can preempt the ongoing task. Therefore, some questions naturally arise: How to select the status update destination, remote cloud server or edge server, to optimize the system performance? Due to the randomness in the actual process, the server-side cannot perform the best action in every time slot; how to allocate the priority between different status updates to get the preemption policy that minimizes the MSE of the system? These questions motivate the study in this paper.

In this paper, the scheduling problem in WNCS with $N$ independent processes is studied, and we are interested in minimizing the average MSE. By referring to the model in \cite{zhou2019minimum} and \cite{han2019ondisc}, we adopt edge computing in the system model and consider that the state updates need to be processed in multiple time slots to generate control commands. Moreover, to handle the status updates more flexibly, we add the preemption mechanism on the basis of \cite{zhou2019minimum}. The main contributions of this paper are summarized as follows:

\begin{itemize}
	\item This paper considers a multi-process WNCS and applies the edge computing. The different processes stands for different physical processes, which status updates need different processing time on the same server. Meanwhile, we use link delay and computing power to describe edge server and cloud server. By introducing the AoI and remaining processing time of each process, the problem of how to scheduling status updates can be described as an infinite horizon average cost Markov Decision Process (MDP). Thus the scheduling policy of minimizing the MSE of the system is obtained. 
	\item A suboptimal policy is proposed to reduce the computational complexity. Along the proof of exists studies, we have the additive separable structure of value function. The original MDP problem can be divided into multiple small tasks. The simulation result shows that the performance of proposed suboptimal policy is near to optimal policy.
	\item The processing preemption mechanism is considered in this paper. In order to handle the status updates more flexibly, we allow the task with higher priority to preempt the processing task on server. The preemption optimal policy is obtained by solving the formulated MDP problem. Then the consistency property of the preemption optimal policy is proved.
\end{itemize}

The rest of this paper is organized as follows. In Section II, we introduce the system model, formulate the problem and prove some structure properties. Section III presents a low-complexity suboptimal policy. Section IV further extends to the preemption mechanism on controller and characterizes the consistency property of its optimal policy. Simulation result and analysis are provided in Section V. Finally, conclusions are drawn in Section VI.
\begin{figure}[h]
	\centering
	\includegraphics[width=0.5\textwidth]{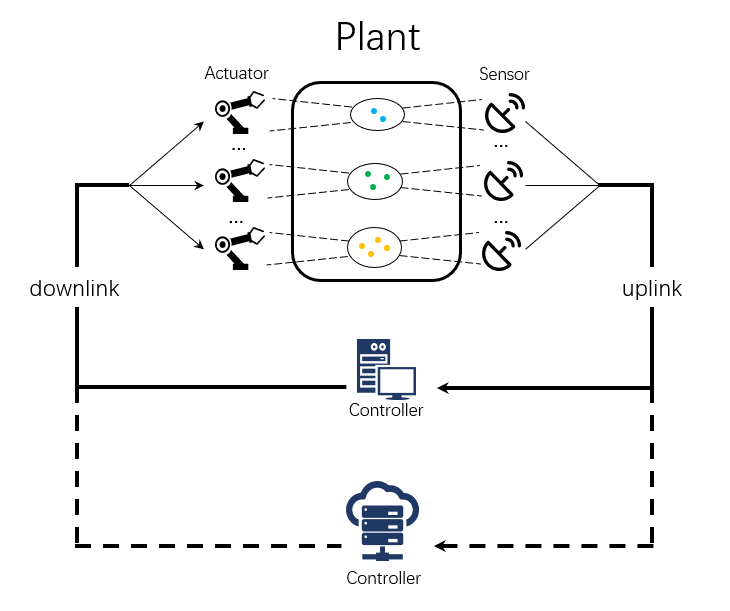}
	\caption{System Model}
\end{figure}

\section{System Model And Problem Formulation}
\subsection{System Model}
We consider a multi-process WNCS. As shown in Fig.1, there are $N$ processes and $N$ sensors in the system, and each sensor samples its corresponding process\cite{feng2018minimizing}-\cite{abd2018average}. For the uplink and downlink wireless channel, we adopt the orthogonal channel which can transmit multiple status updates at the same time. The controller in the system consists of an edge server and a cloud server. Same as the assumption in \cite{han2019ondisc} and \cite{jia2015optimal}, compared with the remote server, the edge server has a smaller uplink and downlink delay but longer task processing time. In Fig.1, different color of dots in the circle represents different physical processes and the number of dots represents the dimension of process, e.g., the processing time required for generating control command from video signals is not the same as that from audio signals.


A time-slotted system is considered, where time is divided into slots with equal duration. And we model each process as a discrete linear time-invariant system \cite{huang2019optimal}:
\begin{equation}
	x_{k}(t+1)=A_kx_k(t)+B_ku_k(t)+z_k(t),
\end{equation}
where $k\in\mathcal{N} \triangleq \{1,... ,N\}$ and $t$ represent the process index and time slot,  respectively. $x_{k}(t)\in R^{n_k}$ is the state of the $k$-th process in $t$-th time slot and $u_k(t)\in R^{n_k}$ is the executed control command.
$z_k(t)$ represents the normally distributed noise with the distribution $N(0,R_k)$ of the $k$-th process and we assume that $R_k$ is a positive semi-definite. $A_k\in R^{n_k\times1}$ and $B_k\in R^{n_k\times1}$ represent the state transition matrix and command control coefficient of process $k$, respectively.
In this article, we consider the case that all processes remain unchanged within a single time slot. The goal of the system is to keep the state $x_k$ of each process $k$ close to $\boldsymbol{0}\in R^{n_k}$.

In wireless channel, the transmission success probability for a status updates is set as $p\in (0,1)$. As described in  \cite{feng2018minimizing} \cite{ceran2018reinforcement}\cite{chen2016age}, it is assumed that there is a perfect feedback channel between each sensor and the destination, so that each sensor will immediately inform whether the transmission is successful and determine the destination of the next time slot.

To generate effective control commands, the controller must maintain an accurate estimate $x_k$ of plant state. When the controller receives the status information from sensor successfully, it can use the timestamp to estimate the plant state. Define $\tau_k$ as the AoI of the status information generated by process $k$. The estimation of $x_k(t)$, denoted by $\widetilde{x}_k(t)$, can be expressed as:
\begin{equation}
	\widetilde{x}_k(t)=A_k^{\tau_k(t)}x_k(t-\tau_k(t)),
\end{equation}


We denote $\Delta^\uparrow$ and $\Delta^\downarrow$ as the uplink and downlink delay, respectively. Assuming that the controller knows the delay of control command transmitted to the actuator. The goal of the system is to keep the state $x_k$ of each process $k$ close to $\boldsymbol{0}\in R^{n_k}$. When the state information has been processed completely, the control command on actuator $u_k(t)$ and the control command on controller $\widetilde{u}_k(t)$ have the following relationship:
\begin{equation}
	u_k(t)=\widetilde{u}_k(t-\Delta^\downarrow)=K_k\widetilde{x}_k(t).
\end{equation}
where $K_k=-A_k/B_k$ is the command generation coefficient.

Let $J$ denote the long-term average MSE of the system, which is given by 
\begin{equation}
	J={\lim_{T \to +\infty}}\frac{1}{T}\sum_{k=1}^{N}\sum_{t=0}^{T}Q_k(t).
\end{equation}
where $Q_k(t)=\mathbb{E}[x_k(t)x_k(t)^T]$ represents the MSE of each process, which can be calculated by combining (1)-(3), given as:
\begin{equation}
	Q_k(t)=\mathbb{E}[x_k(t)x_k(t)^T]=\sum_{i=1}^{\tau_k(t)}(A_k)^{(i-1)}R_k(A_k^T)^{(i-1)}.
\end{equation}

This equation constructs the mapping between AoI and MSE and can simplify the calculation.

For the processing time, different tasks on the same server is not equal, and the same task on different server is also different. For each physical process $k$, we denote the state information length $L_k$, which 
is equal to the number of elements in $x_k$. The number of bits per element in the state matrix and the number of CPU cycles needed to process one bit are denoted by $l$ and $v$, respectively. The number of time slots required for process one status update is then given by \cite{wang2021preprocess}
\begin{equation}
	T_{p}=\left\lceil\frac{L_k l v}{f \varepsilon} \right\rceil.
\end{equation}
where $f$ (in Hz) is the CPU frequency of the processor and $\varepsilon$ (in seconds) is the time duration in each slot.

We assume that the edge computing system is stable such that the link delay is time invariable and a server can execute at most one task at a time. Because the remote cloud servers can be modeled as edge servers but with long link delay and more powerful processing capability, we do not explicitly differentiate between the edge servers and the remote cloud servers\cite{han2019ondisc}. 

In order to describe the control performance of the system more accurately, we make a common assumption in the following:

\emph{\textbf{Assumption:}The transition matrix $A_k$ of each process satisfies that $\rho$($A_k$)>1. It means that without control, the system will be unstable, which is why the control system exists. This assumption makes the control necessary. }

Here $\rho$(.) stands for Spectral radius.


\subsection{Problem Formulation}
For edge servers, the uplink and downlink delays are $\Delta_e^\uparrow$ and $\Delta_e^\downarrow$,  respectively. Moreover, the processing time on edge server of state $x_k$ is defined as $T_{e,k}$. For a cloud server (or called the remote server), the uplink and downlink delays are defined as $\Delta_r^\uparrow$ and $\Delta_r^\downarrow$, respectively. And the processing time on remote server of state $x_k$ is defined as $T_{r,k}$.

In order to facilitate the subsequent representation, we make some assumptions which are similar to the existing works in \cite{han2019ondisc} \cite{liu2020latency}-\cite{gorlatova2018characterizing}. For the edge server, the uplink and downlink delay are set to be $1$ time slot, i.e., $\Delta_e^\uparrow=\Delta_e^\downarrow=1$, and the processing time $T_{e,k}$ is set to be twice the dimension of $x_k$. For the cloud server, we set the uplink and downlink delay to $2$ time slots, i.e., $\Delta_r^\uparrow=\Delta_r^\downarrow=2$, and the processing time $T_{r,k}$ is numerically the same as the dimension of $x_k$. e.g., for a task with a dimension of 1, if dispatched to the edge server, the uplink and downlink delays are both 1, and the processing time is $2$; if dispatched to the cloud server, the uplink and downlink delays are both 2 and the processing time is 1. This can be extended to a general case and will be introduced in the following. But note that, the uploading and downloading of a task do not consume any computing resource of the servers, so a server can process a task while transmitting other tasks at the same time.

Our goal is to jointly control every process to minimize the system MSE under non-uniform status update and edge computing system. This problem can be modeled as a MDP problem, and the system space can be expressed as $\mathbf{S}\triangleq(\tau,C_r,C_e)\in\mathcal{S}$ where $\mathcal{S}$ is the collection of all feasible states. Each parameter is described as follows:

In order to record the AoI information of all processes, we set $\mathbf{\tau}\triangleq(\tau_1,\cdots,\tau_N)$. For different process $k$, we define $\tau_k\triangleq\{1,2,\cdots,\hat{\tau}_k\}$ as the AoI at the beginning of slot $t$. We set $\hat{\tau}_k$ be the upper limits of the AoI for process $k$. For tractability \cite{bertsekas2011dynamic}, we assume that $\hat{\tau}_k$ is finite, but can be arbitrarily large. 

Let $C_r\triangleq(I_{r,1},I_{r,2},I_{r,3},d_r)$ be the state space for the remote server, $I_{r,1}\in\{1,\cdots,N\}$ stands for the index of process which is transmitting in the uplink, $I_{r,2}\in\{1,\cdots,N\}$ stands for the index of process which is being computed on the remote server, $I_{r,3}\in\{1,\cdots,N\}$ stands for the index of process which control command is transmitting from controller. $d_r$ records the number of slot that are left to be computed. 

Let $C_e\triangleq(I_{e,1},d_e)$ be the state space for the edge server, $I_{e,1}\in\{1,\cdots,N\}$ stands for the index of process which is being computed on the edge server, $d_e$ records the number of slot that are left to be computing.

The status information in transmission is like a queue and arrives at the server in the order of FCFS \cite{jia2015optimal}. Therefore, to expand to the general case, just add index element for representing the process according to different uplink and downlink delays.


The action space is defined as $\mathbf{A}\triangleq(a_1,a_2)$, where $a_1,a_2\in\{1,\cdots,N\}$ represent the index of process where status update is transmitted to the remote server and edge server, respectively.

For each variable, we can write the update rules as follows:

When both the uplink and downlink transmission are successful:

\begin{equation}
	\tau_k(t+1)=
	\left\{
	\begin{array}{ll}
		\Delta_r^\uparrow+\Delta_r^\downarrow+T_{r,k}, & \text { if } \ I_{r,3}(t)=k,\\
		\Delta_e^\uparrow+\Delta_e^\downarrow+T_{e,I_{e,1}(t)}, & \text { if }\  d_{e}(t)=0, \\
		\tau_k(t)+1, & \text { otherwise. } 
	\end{array}
	\right.
\end{equation}

\begin{equation}
	\begin{array}{l}
		C_r(t+1)=	\\

		\left\{
		\begin{array}{ll}
			(a_1(t),I_1(t),I_2(t),T_{r,I_1(t)}-1), & \text { if }\  d_r(t)=0, \\
			(a_1(t),I_2(t),0,d_r(t)-1), &\  \text{otherwise.}
		\end{array}
		\right.
	\end{array}
\end{equation}

\begin{equation}
	C_e(t+1)=
	\left\{
	\begin{array}{ll}
		(a_2(t),T_{r,I_1(t)}), & \text { if } \ d_e(t)=1,\\
		(I_{e,1}(t),d_e(t)-1), & \text { otherwise. } 
	\end{array}
	\right.
\end{equation}

In other conditions, (7) will change to $\tau_k(t+1)=\tau_k(t)+1$ when the uplink transmission failed. When the downlink transmission failed, just replace the $a_1(t)$ and $a_2(t)$ to $0$ in (8) and (9), respectively.

The one-stage cost only depends on the current state $s=(\tau,C_r,C_e)$ and is defined as
\begin{equation}
	c(s,a)\triangleq\sum_{k=1}^{N}Q_k=\sum_{k=1}^{N}\sum_{i=1}^{\tau_k}(A_k)^{(i-1)}R_k(A_k^T)^{(i-1)}.
\end{equation}

The set of scheduling decisions for all possible states is called a scheduling policy $\pi\triangleq(a(1),a(2),\cdots)\in\Pi$, where $\Pi$ is the collection of all feasible scheduling policies. As a result, under a feasible stationary policy $\pi$, the average MSE of the system starting from a given initial state $\boldsymbol{S}(1)=\boldsymbol{S}_1$ is given by:
\begin{equation}
	\bar{Q}^{\pi}\left(\boldsymbol{S}_{1}\right) \triangleq \limsup _{T \rightarrow \infty} \frac{1}{T} \sum_{t=1}^{T} \sum_{k=1}^{N} \mathbb{E}\left[Q_{k}(t) \mid \boldsymbol{S}_{1}\right].
\end{equation}

We are seeking for the policy which can minimize average MSE of the system in the following:
\begin{equation} \label{goal}
	\bar{Q}^{*}\left(\boldsymbol{S}_{1}\right) \triangleq \inf _{\pi} \bar{Q}^{\pi}\left(\boldsymbol{S}_{1}\right).
\end{equation}

According to [\citenum{bertsekas2011dynamic}, Propositions 5.2.1], the optimal policy $\pi^*$ can be obtained by solving the following Bellman equation.

\textbf{Lemma 1:} There exists a unique scalar $\theta$ and a value function $\{V(s)\}$ satisfying:

\begin{equation} \label{Bellman}
	{\theta}+{V}(\boldsymbol{s})=c(s,a)+\min_{a\in A}\sum_{s'\in \mathcal{S}}\Pr[s'|s,a]V(s'), \ \forall s \in \mathcal{S},
\end{equation}
where $\theta$ is the optimal value to (\ref{goal}) for all initial state $\boldsymbol{S}_{1}\in\mathcal{S}$ and the optimal policy which can achieve the optimal value $\theta$ satisfies:
\begin{equation} \label{opaction}
	\pi^*(s) = \arg\min_{a\in A}\sum_{s'\in \mathcal{S}}\Pr[s'|s,a]V(s'), \ \forall s \in \mathcal{S}.
\end{equation}

It can be observed that the argument of the value function in (\ref{Bellman}) is only the state $s$. If it is extended to the general situation and the action $a$ is also regarded as the argument of the function, the state-action cost function is obtained as follows:
\begin{equation} \label{stateaction}
	J(\boldsymbol{s},a)=c(s,a)+\sum_{s'\in \mathcal{S}}\Pr[s'|s,a]V(s'), \ \forall s \in \mathcal{S}.
\end{equation}

It is evident from Lemma 1 and (\ref{opaction}) that the optimal policy $\pi^*$ relies upon the value function $V(.)$. In order to obtain $V(.)$, we have to solve the Bellman equation in (\ref{Bellman}). But disappointingly, there is no closed-form solution in (\ref{Bellman}). The numerical solution can only be obtained by iterative methods such as value iteration or policy iteration. Furthermore, the design of the numerical solution corresponding to the optimal policy cannot provide more suggestions and will consume a lot of computing resources, especially in the high dimension problem. 

Therefore, if the structural properties of the optimal policy can be obtained, it can be used to reduce the computational complexity of the policy and verify the final numerical result.

By the dynamics in (7)-(9) and using the relative value iteration algorithm, we can show the following properties of the value function $V(s)$:

\textbf{Theorem 1:} For any state $s^1=(\tau^1,C_r^1,C_e^1)$, $s^2=(\tau^2,C_r^2,C_e^2) $, satisfy that $\tau_1^1\geq \tau_1^2$, $\tau_2^1\geq \tau_2^2$, $\tau_3^1\geq \tau_3^2$, $C_e^2 = C_e^1$ and $C_r^2 = C_r^1$, we have $V(s^1) \geq V(s^2)$.

\emph{Proof:} The proof can be found in the online version.

\textbf{Theorem 2:} For the state $s=(\tau,C_r,C_e)$ which satisfies $C_r=\boldsymbol{0}$ and $C_e=\boldsymbol{0}$ have the following type policy:

If $\pi^*(s)=(k_1,k_2)$, then $\pi^*(s')=(k_1,k_2)$. Where $k_1\neq k_2$ and $s'$ is the same with $s$ except $\tau'_{k_1}=\tau_{k_1}+z$, $\tau'_{k_2}=\tau_{k_2}+z$. $z$ is any positive integer number.

\emph{Proof:} The proof can be found in the online version.

From Theorem 1 and Theorem 2, we present the threshold type policy of the MDP. The structural result will simplify off-line computation and facilitate the online application of scheduling policy. Note that we mainly focus on the case in which the remote server and edge server are idle, i.e., $C_r=\boldsymbol 0$ and $C_e=\boldsymbol 0$. Because the result in other conditions agrees with intuition in other conditions, e.g., when the status information of process $k$ is transmitting to the remote server, or the remote server is computing the status information of process $k$, the edge server will serve other processes except process $k$. Also, this result can be extended to the single server model.

\section{Low-Complexity Suboptimal policy}
In the MDP mentioned above, the computational complexity will become very high when too many elements are in the state space, which is called the curse of dimensionality. So we proposed a suboptimal policy, which sacrifices a small amount of performance in exchange for a significant reduction in computational complexity.

Let $\hat{\theta}$ and $\hat{V}(S)$ be the system average MSE and the value function under an unchained base policy $\hat{\pi}$. From the proof of \cite{zhou2019minimum}, there exists $(\hat{\theta},\hat{V}(\boldsymbol{S}))$ satisfying the following Bellman equation.
\begin{equation} \label{subbellman}
		\hat{\theta}+\hat{V}(\boldsymbol{s})=\min _{a \in A}\left\{c(s, a)+\mathbb{E}\left[V\left(s^{\prime}\right) \mid s, a\right]\right\}, \ \forall s \in \mathcal{S}
\end{equation}
where $s'$ is the next state from state $s$ under the given action $a$. Obviously, the state cost is independent of the action and (\ref{subbellman}) can be further written as
\begin{equation} \label{subrebellman}
	\begin{aligned}
		\hat{\theta}+&\hat{V}(\boldsymbol{s})=\sum_{k=1}^{N} Q_{k} \\
		&+\min_{v}\sum_{s'\in \mathcal{S}}\mathbb{E}^{\hat{\pi}}[\text{Pr}[s'|s,a]]V(s'), \ \forall s \in \mathcal{S}
	\end{aligned}
\end{equation}

For each process $k$, we define $s_k\triangleq(\tau_k,C_r.C_e)\in \mathcal{S}_k$ as the state space and $\mathcal{S}_k$ is the set of all feasible states. The action space is the same as SectionII. The transmit probability can also be derived from the previous description, which is omitted here. After the definition, we show that $\hat{V}(X)$ has the following additive separable structure.

\textbf{Lemma 2:} Given any unchain policy, the value function $\hat{V}(s)$ in (\ref{subbellman}) can be express as $\hat{V}(s)=\sum_{s\in \mathcal{S}}\hat{V}_k(s_k)$, where for each $k$, $\hat{V}_k(s_k)$ have the following property:
\begin{equation} \label{seper}
	\begin{aligned}
		\hat{\theta}+&\hat{V}_k(\boldsymbol{s_k})=Q_{k} \\
		&+\min_{v_k}\sum_{s'_k\in \mathcal{S}_k}\mathbb{E}^{\hat{\pi}}[\text{Pr}[s'_k|s_k,a]]V(s'_k), \ \forall s_k \in \mathcal{S}_k
	\end{aligned}
\end{equation}
where $\theta_k$ and $\hat{V}_k$ are the average MSE and value function of each process under policy $\hat{\pi}$, respectively.

\emph{Proof:} Along the line of proof of [\cite{cui2012delay}, Lemma3], we have the additive separable structure of the value function under a unchain base policy $\hat{\pi}$ and by making use of the relationship between the joint distribution and marginal distribution. So, we have the equation $\sum_{s'\in \mathcal{S}}\text{Pr}[s'|s,a]=\sum_{s'_k\in \mathcal{S}_k}\text{Pr}[s'_k|s,a]=\sum_{s'_k\in \mathcal{S}_k}\text{Pr}[s'_k|s_k,a]$ holds.
Then, by substituting $\hat{V}(s)=\sum_{k\in \mathcal{N}}\hat{V}_k(s_k)$ into (\ref{subbellman}), it can be easily checked that the equality in (\ref{seper}) holds.

Now, we approximate the value function in Bellman equation with $\hat{V}(s):V(s)\approx\hat{V}(s)=\sum_{k\in \mathcal{N}}\hat{V}_k(s_k)$
, and $\hat{V}_k(s_k)$ is given in (\ref{seper}). Then, according to (\ref{subrebellman}) we develop a deterministic scheduling suboptimal policy in the following:
\begin{equation} \label{subpolicy}
	\hat{\pi}^*(s)=\text{arg}\min_{a \in A}\sum_{s'\in \mathcal{S}}\text{Pr}[s'|s,a]\sum_{k\in \mathcal{N}}\hat{V}_k(s'_k),\ \forall s\in \mathcal{S}
\end{equation}

The proposed deterministic policy in (\ref{subpolicy}) likes the one iteration step in the standard policy iteration algorithm. It divides the original MDP into multiple small tasks. Though the number of problem states has not decreased, the dimensionality of each small task has been reduced.

In [\citenum{bertsekas2011dynamic}, proposition 5.4.2] and [\citenum{puterman2014markov}, Theorem 8.6.6], the convergence of the sub-optimal algorithm is proved. In addition, in [\citenum{cui2012delay}, Theorem 1] and [\citenum{zhou2019minimum}, Lemma 3], similar sub-optimal strategies are also adopted, which also proves that the proposed suboptimal algorithm is superior to other random algorithms.

When there are multiple processes in the plant, this suboptimal algorithm will significantly reduce the complexity of the calculation, making it possible to deal with multi-process problems. In general, in order to obtain the deterministic suboptimal policy, it is necessary to calculate the value function of each process $\hat{V}_k(s_k)$, the computational complexity of the proposed sub-optimal algorithm is: $O(\sum_{k\in \mathcal{N}}\hat{A}_k*|C_r|*|C_e|)$. In contrast, the computational complexity required to obtain the optimal policy $\pi$ by calculating $V(s)$ is $O(\prod_{k=1}^N \hat{A}_k*|C_r|*|C_e|)$, where $|C_r|$ and $|C_e|$ represent the number of feasible states in the set $C_r$ and $C_e$, respectively.

\section{Scheduling Under Preemption Mechanism}
Due to the noisy channel, status information cannot transmit successfully in every slot. The controller may be idle if only one status is allowed to send the state to controller at a slot. Moreover, the worse the channel, the longer the controller will be idle. 

The orthogonal channel we adopted allows that multiple sensors can transmit status updates at the same time slot \cite{hosny2019new}, some "standby status information" can be transmitted with the "optimal status information" in the same time slot. When the “optimal status information” transmit failed, the server can process the “standby status information” instead. Work with the preemption mechanism, it can significantly reduce the idle time of the server without reducing the system performance.

In order to facilitate the description of the calculating process on the controller, we call the status update as information task. Preemption means when a higher priority task arrives at the controller, the controller will terminate the currently processing task and execute the newly arrived task. The preempted but not completed task will continue to be executed after the current task is completed when no other preemption occurs later. As shown in Fig.2, task 1 arrives at the server first. When task 1 is being processed, task 2 with higher priority arrives. The server immediately stops executing task 1 and instead performs task 2. When task 3 with a lower priority arrives, the server does not respond. After task 2 is completed, the server continues to execute task 1 that has not been completed before. The preemption mechanism makes that the server can switch to another task and resume the current task later. Therefore, the server can handle tasks more flexibly, and the average MSE of the system will also be reduced.

The goal is the same as Section II, to minimize the system MSE. To investigate the scheduling policy on each controller, we assume the system has $N=2$ sensors, and both sensors can transmit their status updates to server in a time slot using the orthogonal channel. Furthermore, this problem can be extended to general situation. In addition, the task processing time and uplink and downlink delay of the server are the same as the remote server in the system model.

The state space is defined as $\mathbf{S}\triangleq(\mathbf{\tau},\mathbf{T_1},\mathbf{T_2},E_t,I_t)$ which consists all possible states.

$\tau \triangleq (\tau_1,\tau_2)$ which records the AoI of each process. For different process $k$, we defined $\tau_k\triangleq\{1,2,\cdots,\hat{\tau}_k\}$ as the AoI at the beginning of slot $t$. We set $\hat{\tau}_k$ be the upper limits of the AoI for process $k$.

$\mathbf{T_1} \triangleq (I_1,d_1,E_1)$, these parameters characterize the processing task. $I_1\in\{0,1,2\}$ represents the index of the process in which status is computing on controller and $0$ represents no task is processing. $d_1$ records the number of slots that are left to computing. When the task is preempted, the number of time slots in the controller will increase, and $E_1$ records the increasing time slot length of processing tasks.

$\mathbf{T_2}\triangleq (I_2,d_2,E_2)$, these parameters characterize the preempted task. $I_2\in\{0,1,2\}$ represents the index of the preempted task and $0$ represents no task is processing. $d_2$ records the number of slots that are left to computing. $E_2$ records the increasing time slot length of preempted tasks.

\begin{figure}[t]
	\centering
	\includegraphics[width=0.5\textwidth]{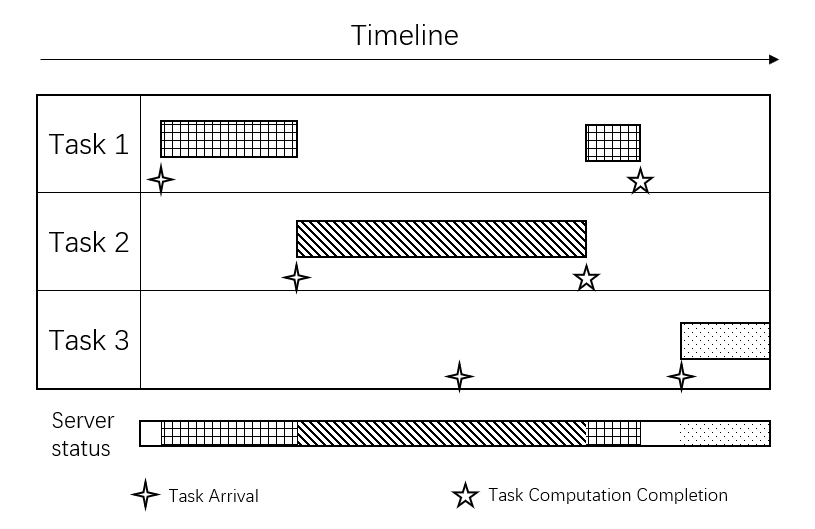}
	\caption{Preemption mechanism on controller.}
\end{figure}

$I_t$ and $E_t$ characterize the task which has been processed completed. $I_t\in\{0,1,2\}$ is the index of the process in which the control command is being transmitted to actuator, and $0$ represents no task is being transmitted. Moreover, $E_t$ records the increasing time slot length of the process in which the control command is being transmitted.

The action $a$ is in the action space $A=\{0,1,2\}$, where $a=0$ means do not preempt and $a=k\ (k\neq 0)$ means the status of process $k$ preempt the processing task. Note that, the current task will be canceled  when $a=k$ and the processing task is from process $k$.Because the current task is obsolete and will not reduced the MSE after completion under this situation.

At the $t$-th time slot, we assume the system state is $s(t)=(\mathbf{\tau}(t),\mathbf{T_1}(t),\mathbf{T_2}(t),E_t(t),I_t(t))$ where  $\mathbf{\tau}(t)=(\tau_1(t),\tau_2(t))$, $T_1(t)=(I_{1}(t),d_1(t),E_1(t))$ and $T_2(t)=(I_{2}(t),d_2(t),E_2(t))$. The state transition formula is given below, and the transition will be written in the order of the success and failure of the downlink channel transmission.

When the downlink transmission is successful, we have:
\begin{equation} \label{tauiter}
	\tau_k(t+1)=
	\left\{
	\begin{array}{ll}
		\Delta_r^\uparrow+\Delta_r^\downarrow+T_{r,k}+E_t(t), & \text { if } \ I_t(t)=k,\\
		\tau_k(t)+1, & \text { otherwise. } 
	\end{array}
	\right.
\end{equation}

\begin{equation} \label{T1iter}
	\begin{array}{l}
		T_1(t+1)=\\

		\left\{
		\begin{array}{ll}
			\{T_{r,k}-1,k,0\}, & \text { if } a=k(k\neq0),\\
			\{d_2(t),I_2(t),E_2(t)\}, & \text { if }a=0,\ d_1=1, \\
			\{\max(d_2(t)-1,0),I_2(t),E_2(t)\}, & \text { if }a=0,\ d_1=0, \\
			\{d_1(t)-1,1(t),E_1(t)\}, & \text { otherwise. } 
		\end{array}
		\right.
	\end{array}
\end{equation}

\begin{equation} \label{T2iter}
	\begin{array}{l}
		T_2(t+1)=\\

		\left\{
		\begin{array}{ll}
			\{d_1(t),I_1,E_1(t)+T_{r,k}\}, & \text { if } a=k\ (k\neq0),\\
			\{0,0,0\}, & \text { if }a=0\ \text{and}\ d_1=1, \\
			\{0,0,0\}, & \text { if }a=0\ \text{and}\ d_1=0, \\
			\{d_2(t),I_2(t),E_2(t)\}, & \text { otherwise. } 
		\end{array}
		\right.
	\end{array}
\end{equation}

\begin{equation} \label{ITiter}
	(I_t(t+1),E_t(t+1))=
	\left\{
	\begin{array}{ll}
		(I_1(t),E_1(t)), & \text { if } \ I_t(t)=k\\
		(0,0), & \text { Otherwise } 
	\end{array}
	\right.
\end{equation}

When the downlink transmission is failed, the only difference with the success situation is in (\ref{tauiter}). The rest part $T_1$, $T_2$, $I_t$ and $E_t$ are the same as (\ref{T1iter}) (\ref{T2iter}) (\ref{ITiter}). Therefore, it will not be described again. Furthermore, the dynamics of the system AoI in (\ref{tauiter}) is changed to:

\begin{equation}
	\tau_k(t+1)=
	\tau_k(t)+1,
\end{equation}

The characteristics of state cost are the same as those in Section II. The one-stage cost only depends on the current state and is defined as:
\begin{equation}
	c(s,a)\triangleq\sum_{k=1}^{N}Q_k=\sum_{k=1}^{N}\sum_{i=1}^{\tau_k}(A_k)^{(i-1)}R_k(A_k^T)^{(i-1)}
\end{equation}

For the preemption model, it has consistency properties.
\textbf{Theorem 3} Consistency. If $a^*(t)=\pi^*(s)=3$, then $a^*(t:t+T_{r,i}-1)=3$ where $a^*(t:t+T_{r,i}-1)=3$ denotes $a^*(t+1)=a^*(t+2)=\cdots=a^*(t+T_{r,i}-1)=3$ and $i$ is the index of status which is calculating on the server.  

\emph{Proof:} See in Appendix C 

In other words, Theorem 3 shows that the process status will not be preempted until it is processed completely when the server takes the optimal action $a^*(t)=\pi^*(s)=i,i=1,2$ successfully.
\section{SIMULATION RESULT AND ANALYSIS}
In this section, we give some parameter settings in simulation and present the numerical result. Besides, the structure of the optimal policies in Section III and IV can verify the numerical result. In addition, we consider a greedy baseline policy for comparison, in which the top two processes with the highest MSE can transmit their status to the controller and the highest one has the priority to choose the controller first. 
\subsection{Setting of system parameters}
In the simulation, we consider four processes and their dimensions decrease in turn:
\begin{equation} \nonumber
\centering {\begin{matrix}
		A_1=\begin{bmatrix}
			1.02 & 1 & 0 & 1\\
			0    & 1 & 0.2&0\\
			0 & 0 & 1 & 0\\
			0 & 0 & 0 & 1.01 
		\end{bmatrix} &   A_2=\begin{bmatrix}
			1 & 1 & 0 \\
			0 & 1.02 &0\\
			0 & 0 & 1
		\end{bmatrix}
\end{matrix}}
\end{equation}

\begin{equation} \nonumber
\centering {\begin{matrix}
		A_3=\begin{bmatrix}
			1.02 & 1\\
			0    & 1
		\end{bmatrix} &   A_4=1.02
\end{matrix}}
\end{equation}
where $A_k$, $k=1,2,3,4$ is the state transmition matrix in (1). As for the noise in (1), like most previous studies, $R_k$ is set to an identity matrix where $k=1,2,3,4$.

The controller parameter settings in the simulation are shown in Table 1. Where the function $size(.)$ can obtain the dimension of independent variable, e.g., $size(A_1)=4$.
\renewcommand\arraystretch{1.5}
\begin{table} \nonumber
	\caption{The controller parameter setting}  
	\begin{center}  
		\begin{tabu} to 0.4\textwidth{X[2,c]|X[3,c]|X[2,c]|X[3,c]}  
			\hline  
			Parameter  &Value             &Parameter     &Value\\  
			\hline  
			$\Delta^\uparrow_r$    &1       &$\Delta^\uparrow_e$           &0   \\  
			$\Delta^\downarrow_r$    &1      &$\Delta^\downarrow_r$        &0 \\  
			$T_{r,k}$    &size($A_k$)      &$T_{e,k}$           &$2*\text{size}(A_k)$\\  
 
			\hline  
		\end{tabu}  
	\end{center}  
\end{table}

\subsection{Scheduling simulation result}
In order to more vividly show the difference in processing tasks between the edge server and the remote server. We first simulate the case where there are only two processes in the system plant.

\begin{figure}[h]
	\centering
	\includegraphics[width=0.5\textwidth]{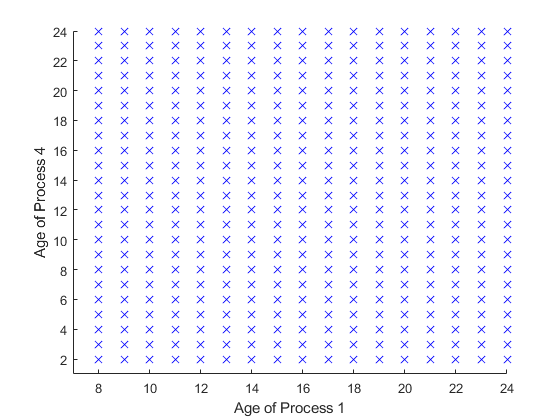}
	\caption{The optimal scheduling policy, where the blue cross stands for action $a =(1,4)$, and $p$ is set to $p=0.9$. Only process1 and process4 in the system.}
	\label{fig3}
\end{figure}

\begin{figure}[h]
	\centering
	\includegraphics[width=0.5\textwidth]{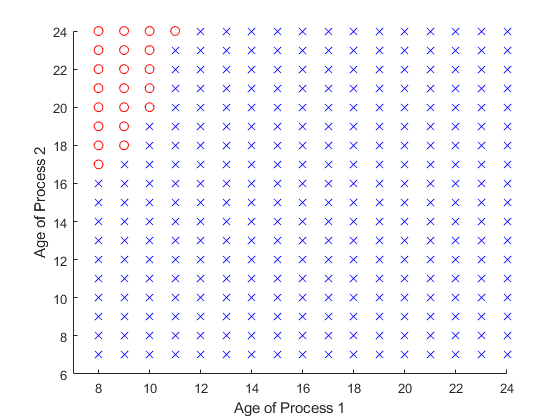}
	\caption{The optimal scheduling policy, where the cross stands for action $a =(1,2)$ and red circle stands for action $a=(2,1)$, and $p$ is set to $p=0.9$. Only process1 and process2 in the system. } \label{fig4}
\end{figure}

In Fig.\ref{fig3} and Fig.\ref{fig4}, the transmission success probability are all set to $p = 0.9$.  Fig.\ref{fig3} is the optimal scheduling policy when only process 1 and process 4 in system. It is obvious that all the actions are $a=(1,4)$, which means the status information from process 1 always send to remote controller and the status information from process 4 always send to edge server. Fig.\ref{fig4} is the optimal scheduling policy when only process 1 and process 2 in system. Different from Fig.\ref{fig3},  there are different actions in different states, and $a^* = (2,1)$ appears in the upper left corner of Fig.\ref{fig4}.

It can be found that there are some subtle differences between the two cases, and the first case is more intuitive, that is, each process corresponds to a controller, e.t., the destination of status updates are deterministic. Nevertheless, in the second case, the optimal policy does not entirely form as two closed loops. It seems a bit counter-intuitive. But under this parameter setting, we can find that, for process 1, if the status is sent to the remote controller, the entire control process needs $\Delta^\uparrow_r+\Delta^\downarrow_r+T_{r,1}=6$ time slots from sampling to completing. Besides, if the status is sent to edge server, the control loop needs $\Delta^\uparrow_e+\Delta^\downarrow_e+T_{e,1}=8$ time slots. For process 2 the control loop needs $\Delta^\uparrow_r+\Delta^\downarrow_r+T_{r,2}=5$ and $\Delta^\uparrow_e+\Delta^\downarrow_e+T_{e,2}=6$, respectively. Obviously, it is the best choice for both process1 and process2 to send the status information to the remote server. Also, the controller can calculate status information when another status is transmitting, i.e., $\Delta^\uparrow_r+\Delta^\downarrow_r =2$ in the control loop can calculate another status. Therefore, when the MSE of process1 is much smaller than the MSE of process2, process 2 will choose a better action for itself, and process 1 can only choose a non-optimal action. However, there is no such trade-off in case 1, where the processing time of status differs significantly. Process1 and process4 are choosing their own optimal conditions, respectively. 
\begin{figure}[h]
	\centering
	\includegraphics[width=0.5\textwidth]{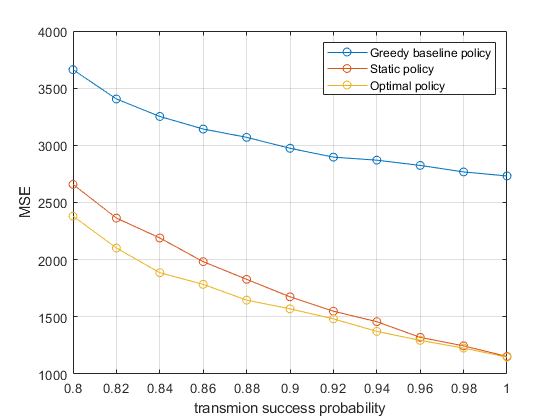}
	\caption{The performance of each policy, when $p=[0.8,1]$. Only process 1 and process 2 in the system.} 
	\label{fig5}
\end{figure}

To verify the performance of various policies, we considered the greedy baseline policy and static policy in addition to the optimal policy. The static policy means that each process will only transmit the status information to the corresponding controller and will not transmit it to other controllers.

 Since the optimal policy in case 1 is entirely equal to the static policy, performance simulation is only performed for case 2. We have done Monte Carlo simulations of 40,000 time slots with different values of $p$. The result is shown in Fig.\ref{fig5}.
 
 Compared with the greedy baseline policy, the performance of optimal policy is greatly improved. While for the static policy, it is vastly improved only when the transmission success probability is low. This is because the optimal policy in Fig.\ref{fig4}, only the upper left corner is different from the static policy, and the lower the transmission success probability, the easier it is to reach these different states. When the transmission success probability is 1, these states will not be reached at all.
 
 For the case where the number of processes in the system is more than that of the controller, we consider the case where process1, process2, and process3 are in the system. When the states of both controllers have been determined, the optimal policy is a three-dimensional diagram. In order to see the optimal policy more intuitively, we intercepted a plane in the three-dimensional coordinate system like Fig.6, which satisfies $\tau_ 1+\tau_ 2+\tau_ 3 = 45 $. 
 
 The results are shown in Fig.7. There are four regions in the graph. It is a threshold policy, which is similar to the optimal policy in a two-dimensional case. When the action is given, the set of actions in the optimal policy is a simply connected region.
%
 
 \begin{figure}[h]
 	\centering
 	\includegraphics[width=0.5\textwidth]{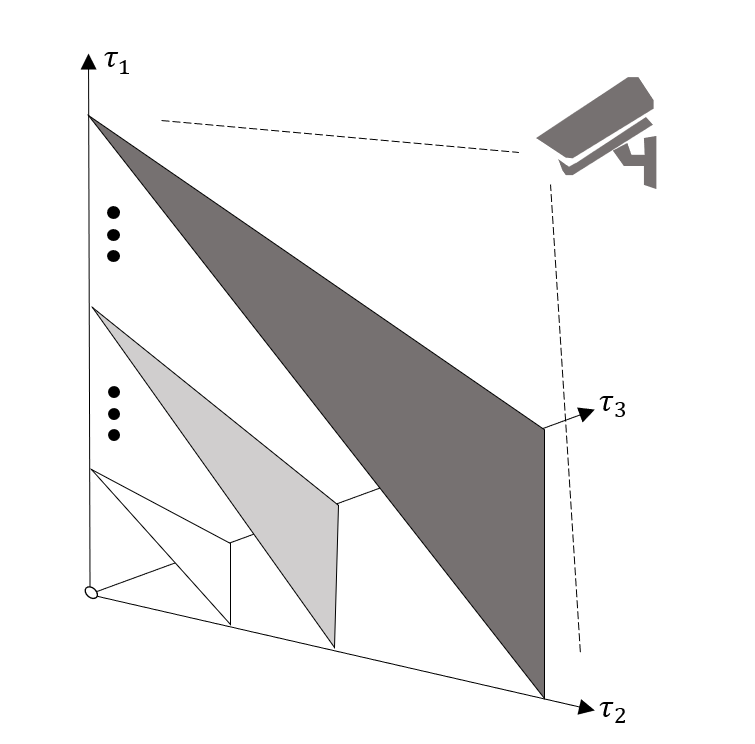}
 	\caption{Schematic diagram of intercepting plane in three-dimensional optimal policy} 
 	\label{fig6}
 \end{figure}

 \begin{figure}[h]
	\centering
	\includegraphics[width=0.4\textwidth]{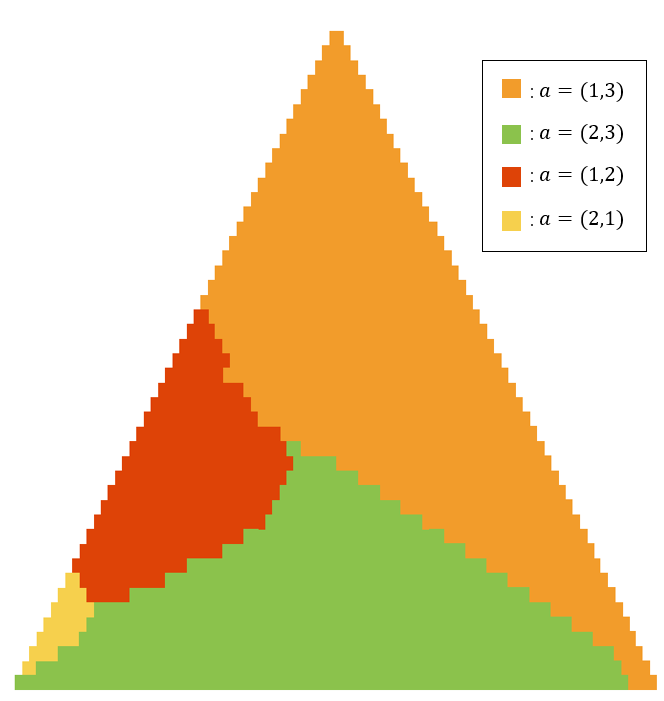}
	\caption{The optimal policy of three processes system} 
	\label{fig7}
\end{figure}

 \begin{figure}[h]
	\centering
	\includegraphics[width=0.5\textwidth]{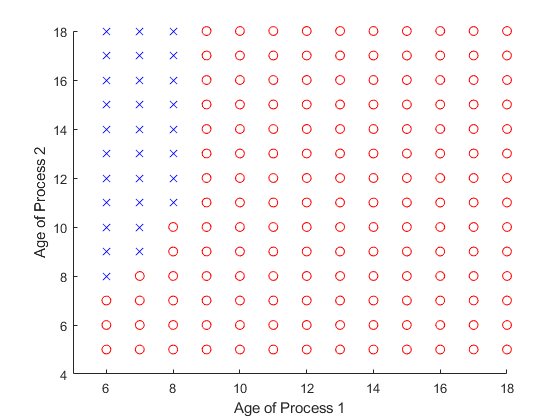}
	\caption{The optimal policy on remote server when edge server is occupied by process 3} 
	\label{fig8}
\end{figure}

 \begin{figure}[h]
	\centering
	\includegraphics[width=0.5\textwidth]{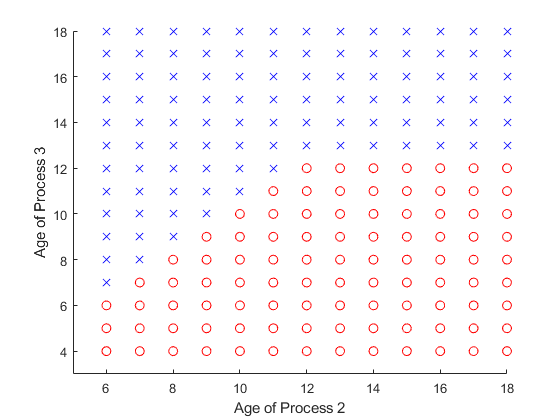}
	\caption{The optimal policy on edge server when remote server is occupied by process 1} 
	\label{fig9}
\end{figure}

Fig.\ref{fig8} represents the optimal policy on remote server when the edge server is occupied by the status from process 3, e.g. $C_r = (I_{r,1},d_{r})=(0,0)$ and $C_e = (I_{e,1},d_{e})=(3,3)$. Fig.\ref{fig9} represents the optimal policy on edge server when the remote server is occupied by the status from process 1, e.g.$C_r = (I_{r,1},d_{r})=(1,4)$ and $C_e =  (I_{e,1},d_{e})=(0,0)$.

 \begin{figure}[h]
	\centering
	\includegraphics[width=0.5\textwidth]{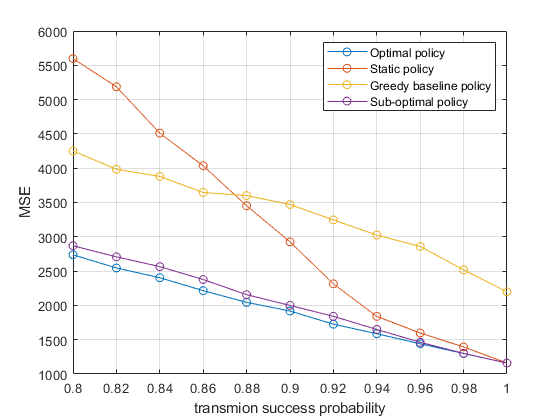}
	\caption{The performance of each policy, when $p=[0.8,1]$.} 
	\label{fig10}
\end{figure}
In Fig.\ref{fig10}, the optimal policy we obtained is compared with sub-optimal, greedy, and static policies. It can be observed that compared with the greedy baseline policy, the proposed optimal policy still has a significant improvement. When the transmission success probability is closer to 1, the performance of optimal policy, sub-optimal policy, and static policy tend to be the same. It is also observed that the performance of sub-optimal policy is close to optimal policy.


\subsection{Application of preemption mechanism}
Fig.11 and Fig.12 are the optimal preemption policy when the remote server processes the tasks of process 1 and process 2, respectively. It is a threshold policy, which is the same as the theory in Section IV. Furthermore, according to the nature of the Theorem 3, when preemption action is executed, the currently running task will not be preempted. 


 \begin{figure}[h]
	\centering
	\includegraphics[width=0.5\textwidth]{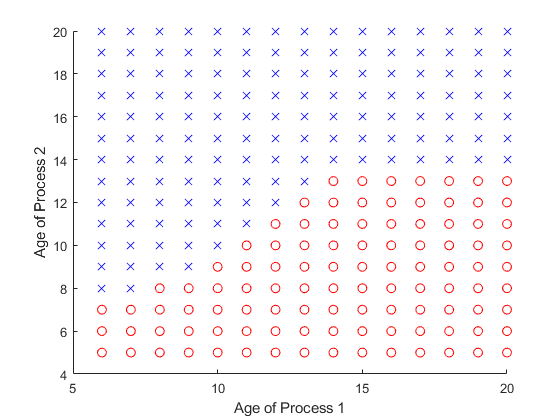}
	\caption{The optimal policy when remote server is calculating the task from process 2. The red circle represents the process 1 task preempts the process 2 task, and the cross represents continue execution of the current task } 
	\label{fig11}
\end{figure}

 \begin{figure}[h]
	\centering
	\includegraphics[width=0.5\textwidth]{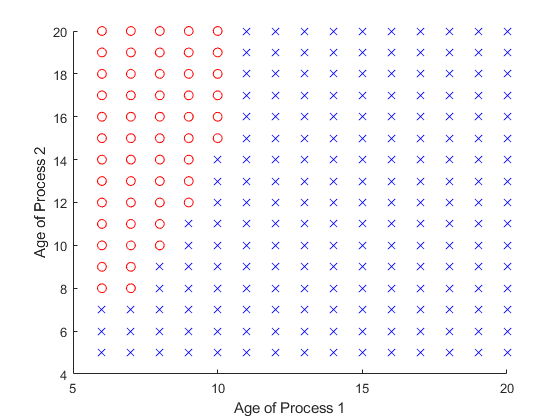}
	\caption{The optimal policy when remote server is calculating the task from process 1. The red circle represents the process 2 task preempts the process 1 task, and the cross represents continue execution of the current task } 
	\label{fig12}
\end{figure}

 \begin{figure}[h]
	\centering
	\includegraphics[width=0.5\textwidth]{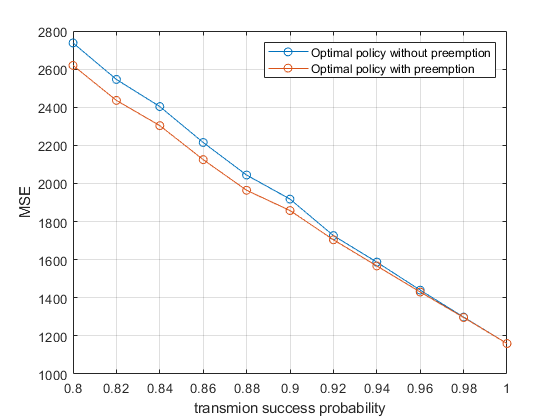}
	\caption{Performance comparison between preemption and no preemption} 
	\label{fig13}
\end{figure}

We compared the performance with the preemption mechanism and the performance without the preemption mechanism. After adding the preemption policy into the optimal policy in Fig.\ref{fig10}, Fig.13 is obtained. It can be found that the higher transmission success probability, the more minor performance improvement, and the overall performance improvement is not significant. Although there are many actions to perform preemption in Fig.11 and Fig.12, preemption is based on the execution of non-optimal actions. Therefore, for WNCS with a periodic status update, although the preemption mechanism will not reduce system performance, it will not significantly improve the system performance.

In the simulation, we can find that compared with the static policy, the performance of our proposed polices is much improved when the transmission success probability is low. Meanwhile, the performance of the optimal policy is the same as that of the static policy when the wireless channel is perfect, i.e., the transmission success probability $p$ is set to $1$. This is because the optimal policy is the same as static policy when the wireless channel is perfect, and the AoI of all processes will change periodically.

\section{Conclusion}
In this paper, we study the scenario of multi-process and edge computing in WNCS intending to minimize the system MSE. We have formulated this problem as an MDP problem and proposed a low complexity suboptimal policy based on random policy. For the obtained optimal policy, by characterizing the monotonicity property of the value function, we prove the structural characteristics of the optimal policy. For the imperfect channel, we also discuss the preemption mechanism on the server and prove that the preemption action is the current optimal action and will not be preempted before the task is completed, which we call consistency properties.

For the preemption mechanism on the server, the performance improvement is not apparent. Because it will not perform non-optimal actions when the channel is reliable. We will further study the event-triggered WNCS system. Because the state generation is random, it is expected that the preemption mechanism will more significantly improve system performance.


%

\appendices
\section{Proof of Theorem 1}
 We prove Lemma 2 using the relative value iteration algorithm(RVIA) [\citenum{bertsekas2011dynamic}, Chapter 5.3] and mathematical induction. In order to make the description clearly, we briefly present the RVIA at the beginning. For each system state $s\in \mathcal{S}$, we denote the value function at iteration $n$ by $V_n(s)$, where $n=1,2,\cdots$. Define the state-action cost function at iteration $n$ as:

\begin{equation} \label{iterJ}
	J_n(s,a)=\sum_{k=1}^{N}Q_k+\sum_{s\in \mathcal{S}}\Pr[s'|s,a]V_n(s').
\end{equation}

Note that $J_n(s,a)$ is related to the right-hand side of the Bellman equation in (\ref{Bellman}). For each $s$, RVIA can be used to find $V_n(s)$ according to:
\begin{equation} \label{iterV}
	V_{n+1}(s)=\min_{a\in A}J_{n+1}(s,a)-\min_{a \in A}J_{n+1}(s^\dagger,a),\ \forall n,
\end{equation}
where $s^\dagger$ is some fixed state. According to [\citenum{bertsekas2011dynamic}, Proposition 5.3.2], the generated sequence $\{V_n(s)\}$ converges to $\{V(s)\}$ under any initialization of $V_0(X)$, i.e.,
\begin{equation} \label{limV}
	\lim_{n\rightarrow +\infty}V_n(s)=V(s),\ \forall s\in \mathcal{S},
\end{equation}
where $V(s)$ satisfies the Bellman equation in (\ref{Bellman}). Let  $\pi^*_n(s)$ be the scheduling action attains to the minimum of the first term in (\ref{iterV}) at the $n$-th iteration for all $s$, i.e.
\begin{equation}
	\pi^*_n(s)=\arg\min_{a \in A}J_{n+1}(s,a),\ \forall s\in \mathcal{S},
\end{equation}
Define $\pi^*_n(s)\triangleq(\pi_{n,k}^*(s))_{k\in \mathcal{N}}$, where $\pi_{n,k}^*(s)$ denotes the scheduling action of the process $k$ under state $s$. We refer to $\pi^*_n$ as the optimal policy at iteration $n$.

By the introduction of above, we will prove Lemma 2 through the RVIA using mathematical induction. Consider two system states $s^1=(\tau^1,C_r^1,C_e^1)$ and $s^2= (\tau^2,C_r^2,C_e^2)$. According to (\ref{limV}), it is sufficient to prove Lemma 2 by show that for any $s^1$ and $s^2$ such that $\tau^2\geq\tau^1$, $C_r^2=C_r^1$ and $C_e^2=C_e^1$. 
\begin{equation} \label{suffL2}
	V_n(s^2)\geq V_n(s^1).
\end{equation}
holds for all $n=1,2,\cdots$

First we initialize $V_1(s)$ for all $s$. We initialize $V_0(s)=0$ and can easily find (\ref{suffL2}) holds for $n=1$. Assume (\ref{suffL2}) holds for some $n>1$. We will prove that (\ref{suffL2}) holds for $n+1$. By (\ref{iterV}), we have
\begin{equation}
	\begin{aligned}
		V_{n+1}\left(\boldsymbol{s}^{1}\right) =&J_{n+1}\left(\boldsymbol{s}^{1}, \pi_{n}^{*}\left(\boldsymbol{s}^{1}\right)\right)-J_{n+1}\left(\boldsymbol{s}^{\dagger}, \pi_{n}^{*}\left(\boldsymbol{s}^{\dagger}\right)\right) \\
		\stackrel{(a)}{\leq}  &J_{n+1}\left(\boldsymbol{s}^{1}, \pi_{n}^{*}\left(\boldsymbol{s}^{2}\right)\right)-J_{n+1}\left(\boldsymbol{s}^{\dagger}, \pi_{n}^{*}\left(\boldsymbol{s}^{\dagger}\right)\right) \\
		= &\sum_{k} Q_{k}^{1}+\sum_{\boldsymbol{s}^{\prime} \in s} \operatorname{Pr}\left[\boldsymbol{s}^{1^{\prime}} \mid \boldsymbol{s}^{1}, \pi_{n}^{*}\left(\boldsymbol{s}^{2}\right)\right] V\left(\boldsymbol{s}^{1^{\prime}}\right) \\
		&-J_{n+1}\left(\boldsymbol{s}^{\dagger}, \pi_{n}^{*}\left(\boldsymbol{s}^{\dagger}\right)\right),
	\end{aligned}
\end{equation}
where $(a)$ is due to the optimal of $\pi_{n}^*(s^1)$ for $s^1$ at iteration $n$. By (\ref{iterJ}) and (\ref{iterV}), we have 
\begin{equation}
	\begin{aligned}
		V_{n+1}\left(\boldsymbol{s}^{2}\right)=& J_{n+1}\left(\boldsymbol{s}^{2}, \pi_{n}^{*}\left(\boldsymbol{s}^{2}\right)\right)-J_{n+1}\left(\boldsymbol{s}^{\dagger}, \pi_{n}^{*}\left(\boldsymbol{s}^{\dagger}\right)\right) \\
		=& \sum_{k} Q_{k}^{2}+\sum_{\boldsymbol{s}^{\prime} \in \mathcal{s}} \operatorname{Pr}\left[\boldsymbol{s}^{2^{\prime}} \mid \boldsymbol{s}^{2}, \pi_{n}^{*}\left(\boldsymbol{s}^{2}\right)\right] V\left(\boldsymbol{s}^{2^{\prime}}\right) \\
		&-J_{n+1}\left(\boldsymbol{s}^{\dagger}, \pi_{n}^{*}\left(\boldsymbol{s}^{\dagger}\right)\right).
	\end{aligned}
\end{equation}

Then we compare $\sum_{\boldsymbol{s}^{1^{\prime}} \in \mathcal{S}} \operatorname{Pr}[{s}^{{1}'} | \boldsymbol{s}^{1}, \pi_{n}^{*}(\boldsymbol{s}^{2})] V(\boldsymbol{s}^{\mathrm{l}'})$ with $\sum_{\boldsymbol{s}^{2^{\prime}} \in \mathcal{S}} \operatorname{Pr}[{s}^{{2}'} |\boldsymbol{s}^{2}, \pi_{n}^{*}(\boldsymbol{s}^{2})] V(\boldsymbol{s}^{\mathrm{l}'})$ for all possible $\pi_{n}^{*}\left(\boldsymbol{s}^{2}\right)=(\pi_{n, k}^{*}(\boldsymbol{s}^{2}))_{k \in \mathcal{N}}$. For each process $k$, we need to consider all cases under the feasible actions. According to (7)-(9), we can check that $Q_k^{2'} \geq Q_k^{1'}$, $C_r^{2'}=C_r^{1'}$ and $C_e^{2'}=C_e^{1'}$ hold for all corresponding actions. Thus, by the induction hypothesis, we have $\sum_{\boldsymbol{s}^{1^{\prime}} \in \mathcal{S}} \operatorname{Pr}[{s}^{{1}'} | \boldsymbol{s}^{1}, \pi_{n}^{*}(\boldsymbol{s}^{2})] V(\boldsymbol{s}^{\mathrm{l}'}) \geq \sum_{\boldsymbol{s}^{2^{\prime}} \in \mathcal{S}} \operatorname{Pr}[{s}^{{2}'} |\boldsymbol{s}^{2}, \pi_{n}^{*}(\boldsymbol{s}^{2})] V(\boldsymbol{s}^{\mathrm{l}'})$ which implies that $V_{n+1}(s^2) \geq V_{n+1}(s^1)$, i.e., (\ref{suffL2}) holds for $n+1$. Therefore, by induction, we know that (\ref{suffL2}) holds for any $n$. By taking limits on both sides of (\ref{suffL2}) and by (\ref{limV}). Now, we complete the proof.

\section{Proof of Theorem 2}
From \cite{wu2017optimalschduling} and \cite{ren2018attack} we define the discounted cost under a deterministic and stationary policy $\pi$ and an initial state $s_0$ by
\begin{equation} \label{policyvalue}
	J_{\alpha}\left(s_{0}, \pi\right)=\limsup _{T \rightarrow \infty} \frac{1}{T} \mathbb{E}_{s}^{\pi}\left[\sum_{k=0}^{T-1} \alpha^{k} c\left(s_{k}, a_{k}\right)\right],
\end{equation}
Note that (\ref{policyvalue}) is different from (\ref{stateaction}), the independent variable is policy instead of action. Define $J_\alpha^*(s_0)=\inf_{\pi \in \Pi}J_\alpha(s_0,\pi)$ and $u(s)=J_\alpha^*(s)-J_\alpha^*(s^\dagger)$, where $s^\dagger$ is some fixed state. Because there exists an optimal policy to the average cost counterpart of the cost function (\ref{policyvalue}), the limit of $u(s)$ as $\aleph$ goes to $1$ exits and is the relative value function in (\ref{Bellman}), i.e.,
\begin{equation} \label{map}
	V(s)=\lim_{\alpha \rightarrow 1}u(s),
\end{equation}

From the mapping in (\ref{map}), we can analyze the properties of $V(s)$ by examining $u(s)$ by value iteration. We further define the dynamic programming operator $T_\alpha u(.)$ for a given measurable function $u$: $\mathcal{S} \mapsto \mathcal{R}$ as
\begin{equation}
	T_\alpha u(s) \triangleq \min_{a \in A}c(s,a)+\alpha\mathbb{E}_s^\pi[u],\ s\in \mathcal{S},
\end{equation}
$T_\alpha u(s)$ is a contraction mapping and follows from Hern$\acute{a}$ndez-Lerma [24]. By Banach fixed-point theorem,
\begin{equation} \label{guarantee}
	\lim_{n \rightarrow \infty} T_\alpha^n u(s)=J_\alpha^*(s).
\end{equation}
If the dynamic operator preserves such properties, (\ref{guarantee}) can guarantees that certain properties holds for $u(s)$. e.g. we can obtain the structure of $V(s)=\lim_{\alpha \rightarrow 1}u(s)$ by verifying that the dynamic operator (\ref{map}) preserves the same structure[14].

For the clarity of later proof, we set the number of processes in the plant to 3 and express the state space as $s=\left( \tau, C_r, C_e\right) =\left((\tau_1,\tau_2,\tau_3),C_r, C_e \right) $. 
In addition, the value of $c(s,a)$ is only related to $\tau$, so we set $c(s,a)=c(\tau)$ in the following proof. 

Take action $a^*=(1,2)$ as the optimal action of state $s$, and compare it with action $a=(2,3)$ here. The other non-optimal actions can also be compared with the same steps. This Theorem is equivalent to the following:

\noindent (1) If $V\left( \tau,C_r^{1},C_e^{2}\right) \leq V\left( \tau,C_r^{2},C_e^{3}\right)$, then $V\left( \tau',C_r^{1},C_e^{2}\right) \leq V\left( \tau',C_r^{2},C_e^{3}\right)$. 

\noindent (2) If $V\left( \tau,C_r^{1},C_e^{2}\right) \leq V\left( \tau,C_r^{2},C_e^{3}\right)$, then $V\left( \tau',C_r^{1},C_e^{2}\right) \leq V\left( \tau',C_r^{2},C_e^{3}\right)$. 

\noindent where $C_r^1=(1,0,0,0)$, $C_e^2=(T_{e,2}-1,2)$, $C_r^2=(2,0,0,0)$, $C_e^3=(T_{e,3}-1,3)$ and $\tau'=(\tau_1+z,\tau_2+z,\tau_3)$, $z$ is any positive integer.

Then we prove (1) the structure properties by showing that $u(.)$ has the same structure, and (2) can be proved in the same way.

In the case that $T_{e,2}> 1$ and $T_{e,3} > 1$.

If $u\left( \tau,C_r^{1},C_e^{2}\right) \leq u\left( \tau,C_r^{2},C_e^{3}\right)$, from Theorem 2, we obtain

\begin{equation} \label{fake}
	\begin{aligned}
		c(\tau_1,\tau_2,\tau_3)+&u(\tau+\boldsymbol{1},C_r^{1},C_e^{2}) \\
		&\leq c(\tau_1,\tau_2,\tau_3)+u(\tau+\boldsymbol{1},C_r^{2},C_e^{3}),
	\end{aligned}
\end{equation}
where $\tau+\boldsymbol{1}=(\tau_1+1,\tau_2+1,\tau_3+1)$, and (\ref{fake}) implies that 
\begin{equation} \label{faker}
	u(\tau+\boldsymbol{1},C_r^{1},C_e^{2}) \leq u(\tau+\boldsymbol{1},C_r^{2},C_e^{3}),
\end{equation}

Then we assume $u\left( \tau',C_r^{1},C_e^{2}\right) > u\left( \tau',C_r^{2},C_e^{3}\right)$, the same with (\ref{fake}) we have
\begin{equation} \label{fake1}
	\begin{aligned}
		u( \tau'&,C_r^{1},C_e^{2}) \\
		&> c(\tau_1+z,\tau_2+z,\tau_3) + u(\tau'+\boldsymbol{1},C_r^{2},C_e^{3}),
	\end{aligned}
\end{equation}
beacuse $c(\tau_1+z,\tau_2+z,\tau_3)>0$ which implies that
\begin{equation} \label{fake2}
	u( \tau',C_r^{1},C_e^{2})>u(\tau'+\boldsymbol{1},C_r^{2},C_e^{3}),
\end{equation} 
Meanwhile, from Lemma 2 there exist
\begin{equation} \label{fake3}
	u(\tau'+\boldsymbol{1},C_r^{2},C_e^{3})>u(\tau+\boldsymbol{1},C_r^{2},C_e^{3}),
\end{equation} 
From (\ref{fake1})-(\ref{fake3}) we have
\begin{equation} 
	u( \tau',C_r^{1},C_e^{2})>u(\tau+\boldsymbol{1},C_r^{2},C_e^{3}),
\end{equation} 
According to the existing conditions (\ref{faker})
\begin{equation} 
	u(\tau+\boldsymbol{1},C_r^{2},C_e^{3})  \geq  u(\tau+\boldsymbol{1},C_r^{1},C_e^{2}),
\end{equation}
which causes 
\begin{equation} \label{vio}
	u( \tau',C_r^{1},C_e^{2})  >  u(\tau+\boldsymbol{1},C_r^{1},C_e^{2}).
\end{equation}

When $z=1$ (\ref{vio}) violate the monotonicity and consistency of the value function and hence the assumption is incorrect. Therefore,  $u\left( \tau',C_r^{1},C_e^{2}\right) \leq u\left( \tau',C_r^{2},C_e^{3}\right)$ when $z=1$. Same as the proof in Lemma2, it can be obtained by mathematical induction that this formula holds for any positive integer $z$. The proof is done and other cases can be proved in the same steps.

\section{Proof of theorem 3}
Consider the case that $T_{r,1}>2$ and $T_{r,2}>2$. Other cases can be easily extended. In order to simplify the expression, we omit the completely equal parameters in the proof. The state can be expressed as $s=(\tau_1,\tau_2,d_1,d_2)$, where $d_1$ and $d_2$ represent the remaining computing time of the two processes in plant, respectively. Without loss of generality, assume that $a^*(t-1)=1,\ a^*(t+1)=3$, i.e.the server will compute the status from sensor 1 at $(t-1)$-th time slots, and will continue calculate the status information of the last time slot at $t-$th time slot.

By the assumption, we have
\begin{equation} \label{consist1}
	\begin{aligned}
		c(\tau_1,\tau_2)&+u(\tau_1+1,\tau_2+1,T_{r,1}-2,T_{r,2})\\
		&<c(\tau_1,\tau_2)+u(\tau_1+1,\tau_2+1,T_{r,1}-1,T_{r,2}),
	\end{aligned}
\end{equation}

\begin{equation} \label{consist2}
	\begin{aligned}
		c(\tau_1,\tau_2)&+u(\tau_1+1,\tau_2+1,T_{r,1}-2,T_{r,2})\\
		&<c(\tau_1,\tau_2)+u(\tau_1+1,\tau_2+1,T_{r,1}-1,T_{r,2}-1),
	\end{aligned}
\end{equation}

(\ref{consist1}) indicates the action $a(t)=3$ is better than $a(t)=1$, and (\ref{consist2}) indicates the action $a(t)=3$ is better than $a(t)=2$. Meanwhile, the inequality (\ref{consist1}) and (\ref{consist2}) also can be further simplified to:

\begin{equation} \label{consist3}
	u(\tau_1+1,\tau_2+1,T_{r,1}-2,T_{r,2})
	<u(\tau_1+1,\tau_2+1,T_{r,1}-1,T_{r,2}),
\end{equation}

\begin{equation} \label{consist4}
	u(\tau_1+1,\tau_2+1,T_{r,1}-2,T_{r,2})
	<u(\tau_1+1,\tau_2+1,T_{r,1}-1,T_{r,2}-1),
\end{equation}
If $a^*(t+1)=1$, then the Left side of inequality (\ref{consist1}) can be expressed as
\begin{equation} \label{consist5}
	\begin{aligned}
		c(\tau_1,\tau_2)+c(\tau_1+1,\tau_2+1)+u(\tau_1+2,\tau_2+2,T_{r,1}-1,T_{r,2}),
	\end{aligned}
\end{equation}
Because $c(\tau_1+1,\tau_2+1)>0$, so we have:
\begin{equation} \label{consist6}
	u(\tau_1+1,\tau_2+1,T_{r,1}-2,T_{r,2})>u(\tau_1+2,\tau_2+2,T_{r,1}-1,T_{r,2}),
\end{equation}
One the other hands, due to the monotonicity, we have
\begin{equation} \label{consist7}
	u(\tau_1+2,\tau_2+2,T_{r,1}-1,T_{r,2})>u(\tau_1+1,\tau_2+1,T_{r,1}-1,T_{r,2}),
\end{equation}
Combine (\ref{consist6}) and (\ref{consist7}), it is obviously that
\begin{equation} \label{consist8}
	u(\tau_1+2,\tau_2+2,T_{r,1}-2,T_{r,2})>u(\tau_1+1,\tau_2+1,T_{r,1}-1,T_{r,2}).
\end{equation}

We can found that (\ref{consist8}) is contradict to (\ref{consist3}). Therefore, different from proving that action $a^*(t+1)=3$ is optimal, the contradiction just prove $a(t+1)=3$ is better than $a(t+1)=1$. It is also necessary to prove that action $a(t+1)=3$ is better than the rest. The proof is exactly the same as (\ref{consist1})-(\ref{consist8}), and is omitted here.





\bibliographystyle{ieeetr}
\bibliography{bare_jrnl}
\end{document}